\documentclass{article}
\usepackage{amsfonts}
\usepackage{amsmath}

\newtheorem{theorem}{Theorem}

\newenvironment{proof}[1][Proof]{\noindent\textbf{#1.} }{\ \rule{0.5em}{0.5em}}
\input{tcilatex}

\begin{document}

\begin{center}
{\huge New Cellular Automata associated with the Schroedinger Discrete
Spectral Problem}

\bigskip

\textbf{M. Bruschi}$^{\ast }$\ 

Dipartimento di Fisica, Universit\`{a} di Roma "La Sapienza", 00185 Roma,
Italy

Istituto Nazionale di Fisica Nucleare, Sezione di Roma

\textit{E-mail: Mario.Bruschi@roma1.infn.it}

\bigskip \bigskip

\textit{Abstract}
\end{center}

New Cellular Automata associated with the Schroedinger discrete spectral
problem are derived. These Cellular Automata possess an infinite (countable)
set of constants of motion.

\bigskip

\bigskip \bigskip

\bigskip -------------------------

{\small Mathematics Subject Classification: }37B15, 35Q51

{\small Physics and Astronomy Classification Scheme: 02.30.Ik }

{\small \bigskip }\newpage

\section{Introduction}

\bigskip A few years ago a class of 1-dimensional "filter" Cellular Automata
(CA) \cite{wolfram},\cite{filter} associated with the Schroedinger discrete
spectral problem (SDSP) was introduced \cite{noi}. These CA are endowed with
a number of interesting properties, namely:

\begin{itemize}
\item they possess an infinite (countable) set of constants of motion

\item they are time reversible

\item they exhibit an interesting dynamic (solitons)
\end{itemize}

However the authors of \cite{noi} were not able to fully develop the usual
integration scheme (solving the direct and inverse spectral problem) for
these CA; this is due to two technical reasons:

\begin{enumerate}
\item these CA are "filter" CA \cite{filter}: i.e. their evolved state at a
given time $t$ at a given lattice position $n$ depends not only on the state
of these CA at time $\ t-1$ but also on the "evolved" state of the CA
themselves \ at the positions $\ j<n$; thus, although these CA are
computable, their evolution is not given in an explicit form;

\item it was impossible to find a Jost solution of the related SDSP at $%
-\infty $, due to the difficulty of projecting rational expressions on a
finite field.
\end{enumerate}

In this paper we introduce three new CA associated to SDSP, and therefore
endowed with an infinite (countable) set of constants of motion, which
(separately) overcome these two difficulties: indeed the first one (CA1) is
a standard CA (not a "filter" one) and the second and the third ones (CA2a,
CA2b) are direct projections of the Backlund Transformation (BT) of the
Kac-van Moerbeke Lattice \cite{kac} on a finite field. This very natural way
to obtain a CA associated to a discrete spectral problem, namely to project
the related BT on a finite field, is usually not exploited, due to the
presence of denominators in the BT itself: in this paper \ we show how this
difficulty can be overcome, by selecting properly the modulus of the finite
field and the parameterization of the fields to be projected. We believe
that these results are important as a necessary (if preliminary) step in
order to develop a fully integration scheme in finite fields (work is in
progress in this direction).

\section{Spectral problem, associated CA and their constants of motion}

First we show how to construct CA from the Schroedinger discrete spectral
problem, namely:

\begin{subequations}
\label{qw}
\begin{equation}
\psi (n-1,z)+u(n)\psi (n+1,z)=(z+z^{-1})\psi (n,z);  \label{ds1}
\end{equation}%
where%
\begin{equation}
n\in 
%TCIMACRO{\U{2124} }%
%BeginExpansion
\mathbb{Z}
%EndExpansion
;~z\in 
%TCIMACRO{\U{2102} }%
%BeginExpansion
\mathbb{C}
%EndExpansion
;  \label{ds2}
\end{equation}%
\begin{equation}
u(n)\underset{\left\vert n\right\vert \longrightarrow \infty }{%
\longrightarrow }1~.  \label{ds3}
\end{equation}

Assuming that the eigenfunction $\psi $ as well as the field $u$ depend also
parametrically on the the discrete time variable $t$ ($t\in 
%TCIMACRO{\U{2124} }%
%BeginExpansion
\mathbb{Z}
%EndExpansion
)$, we look for a compatible evolution of $\psi $ in the form: 
\end{subequations}
\begin{equation}
\psi (n,z,t+1)=\dsum\limits_{j=0}^{M}U^{\left( j\right) }\left( n,t\right)
\psi (n+2j,z,t)~.  \label{se}
\end{equation}

\bigskip

\bigskip Even if the results exhibited in the following could be easily
extended to the general case, for the sake of simplicity we will restrict
considerations only to the case $M=1,U^{\left( 0\right) }=1,U^{\left(
1\right) }=U$ ; in this simple case compatibility between (\ref{qw}) and (%
\ref{se}) yields: 
\begin{subequations}
\label{cc}
\begin{equation}
u(n,t+1)+U(n-1,t)=u(n,t)+U(n,t)~,  \label{c1}
\end{equation}%
\begin{equation}
u(n,t+1)U(n+1,t)=u(n+2,t)U(n,t)~.  \label{c2}
\end{equation}

The ordinary solution of (\ref{cc}) obtains solving f.i. eq. (\ref{c1} ) in
terms of $U\ $and then inserting the obtained expression of $U$ in eq. (\ref%
{c2}), or (equivalently, even if the explicit solutions looks quite
different), solving first eq. (\ref{c2}) and then eq. (\ref{c1}). These two
solutions are the simplest Backlund Transformations of the Kac-van Moerbeke
lattice \cite{kac} : 
\end{subequations}
\begin{subequations}
\label{ca2}
\begin{equation}
u(n,t+1)=u(n,t)+\left( \frac{u(n+1,t)}{u(n-1,t+1)}-1\right) \left(
b+\dsum\limits_{j=-\infty }^{n-1}\left( u(j,t+1)-u(j,t\right) \right) ,
\label{Ca2a}
\end{equation}%
\begin{equation}
u(n,t+1)=u(n,t)+b\left( \frac{u(n+1,t)}{u(n-1,t+1)}-1\right)
\dprod\limits_{j=-\infty }^{n-1}\frac{u(j+1,t)}{u(j-1,t+1)}.  \label{ca2b}
\end{equation}%
where $b$ is an arbitrary constant (the BT parameter).

In order to view eqs. (\ref{ca2}) as defining a CA, one should project them
on a finite ring, say $Z_{m}=Z/m=\{0,1,..,m-1\}$, thus the problem to
project possibly null denominators arises. In \cite{noi} a strategy to
overcome this difficulty was developed: eqs. (\ref{cc}) were considered as "
modulo congruences" an in particular eq.(\ref{c1}) was solved in terms of $\
U$ in the finite ring $\ Z_{m}$ equating each member of the equation itself
to zero. Such peculiar solution reads (here and in the following the symbol $%
\overset{m}{=}$ denotes a $m$-modulo congruence): 
\end{subequations}
\begin{equation}
U(n,t)\overset{m}{=}h~\delta \left( u(n-1,t+1)\right) \delta \left(
u(n+2,t)\right) ,  \label{s1}
\end{equation}

where $\ h$ , although assumed to be a constant in the following, could be
in principle an arbitrary function of the field $u$, while the modulo\_delta
function is defined as:%
\begin{equation}
\delta (x)\overset{m}{=}0~~iff~x\neq 0~\func{mod}~m,~x\in Z_{m}.
\label{del1}
\end{equation}%
It admits the following simple representation:%
\begin{equation}
\delta (x)\overset{m}{=}d\dprod\limits_{k=1}^{m-1}(x+k),~  \label{del2}
\end{equation}%
where $d$ is a normalization constant; note that, if the modulus $m$ is a
prime number, then%
\begin{equation}
d=m-1\Longrightarrow \delta (0)\overset{m}{=}1.
\end{equation}%
Inserting (\ref{s1}) into (\ref{c1}) an interesting "filter" CA obtains
(CA0):%
\begin{eqnarray}
&&u(n,t+1)\overset{m}{=}u(n,t)+h~\delta \left( u(n-1,t+1)\right) \delta
\left( u(n+2,t)\right) +  \notag \\
&&+h(m-1)~\delta \left( u(n-2,t+1)\right) \delta \left( u(n+1,t)\right) ,~
\end{eqnarray}%
whose properties were investigated in \cite{noi}. However the " filter" and
thus rather implicit nature of CA0 is disturbing in order to fully
investigate the mathematical properties of the CA itself.

As a first result of this paper we show that, within the same strategy
adopted in \cite{noi}, there is an alternative solution of eqs. (\ref{cc})
that allows to construct a new CA (CA1) which is "standard", i.e. a non
"filter" one. Indeed inserting (\ref{c1}) into (\ref{c2}), one gets:%
\begin{equation}
\left( u(n,t)+U(n,t)-U(n-1,t)\right) U(n+1,t)=u(n+2,t)U(n,t).
\end{equation}%
It is straightforward to see that the above equation admits, in the finite
ring $Z_{m}$, the following solution:%
\begin{equation}
U(n,t)\overset{m}{=}h~u(n,t)u(n+1,t)\delta \left( u(n-1,t)\right) \delta
\left( u(n+2,t)\right) ,~  \label{sol1}
\end{equation}

where $\delta (x)\ $\ is given by (\ref{del1},\ref{del2}) and $h$ is again,
in principle, an arbitrary function of the field (however, for the sake of
simplicity, we will assume it to be a constant). Inserting (\ref{sol1}) in (%
\ref{c1}) , the following "non filter" CA obtains (CA1):%
\begin{equation*}
u(n,t+1)\overset{m}{=}u(n,t)+h~u(n,t)u(n+1,t)\delta \left( u(n-1,t)\right)
\delta \left( u(n+2,t)\right) +
\end{equation*}%
\begin{equation}
+h~(m-1)u(n-1,t)u(n,t)\delta \left( u(n-2,t)\right) \delta \left(
u(n+1,t)\right) .
\end{equation}%
Note that CA1 loses the symmetry%
\begin{equation}
u(n+j,t)\leftrightarrow u(n-j,t+1),
\end{equation}%
which implied the "time reversibility of CA0 (see \cite{noi}); however CA1
is parity-invariant and moreover it possess a countable set of constants of
motion (see below). For the sake of completeness, let us exhibit the
explicit form of CA1 in the simplest non trivial case ($h=1,~m=2$):%
\begin{equation*}
u(n,t+1)\overset{2}{=}u(n,t)+u(n,t)u(n+1,t)\left( 1+u(n-1,t)\right) \left(
1+u(n+2,t)\right) +
\end{equation*}%
\begin{equation}
+u(n-1,t)u(n,t)\left( 1+u(n-2,t)\right) \left( 1+u(n+1,t)\right) .
\end{equation}%
Now, as a second result of this paper, we show how to project directly the
BT (\ref{ca2}) on a finite field $Z_{m}$ obtaining two new "filter" automata
(CA2a, CA2b). In our opinion this direct projection should preserve all the
properties of the integrability scheme (\ref{qw},\ref{se}). To do so, we
need the following \emph{Theorem }(it should be well known to the experts in
number-theory, however, for the sake of completeness and for the convenience
of the reader, we give also a neat proof of the theorem itself).\medskip

\begin{theorem}
Consider the finite ring $Z_{m}=Z/m=\{0,1,..,m-1\}$ with:%
\begin{equation}
m=p^{k},~k\in 
%TCIMACRO{\U{2115} }%
%BeginExpansion
\mathbb{N}
%EndExpansion
,~p=prime.  \label{p1}
\end{equation}%
Consider also the set $Y\subset $ $Z_{m}$ of all the elements:%
\begin{equation}
y_{s}\overset{m}{=}s\cdot ~p~,~s=0,1,...  \label{p2}
\end{equation}%
Then:\linebreak \newline
1) $Y$ is a (sub)ring ($\func{mod}~m);$\newline
\newline
2) its complementary $X=Z_{m}-Y~$is a group;\newline
\newline
3) $\left( x+y\cdot z\right) \in X,~$for any $x\in X,~y\in Y,~z\in
Z_{m}.\smallskip $
\end{theorem}

\begin{proof}
1) It is evident that $Y$ is closed with respect to addition and
multiplication, see (\ref{p2}).\newline

2) Obviously $1\in X$ (see again (\ref{p2})). Moreover we know from
elementary number theory that the modulo congruence $a\cdot z\overset{m}{=}%
c~ $ admits exactly $\left( a,m\right) $ different solutions $z\in Z_{m}~$ $%
iff$ \ $\left( a,m\right) $ divides $c$ (here and in the following $\left(
a,b\right) $ denotes the \emph{h.f.c.} of $\ a$ and $b$). Since obviously $%
\left( x,m\right) =1$ for any $x\in X$, it follows that any element of $X$ \
has a unique inverse in $Z_{m}$. On the contrary, no element of $Y$ can have
an inverse ($\left( y,m\right) =\left( s\cdot p,p^{k}\right) \neq 1$): thus,
since $X\cap Y=0,~X\cup Y=Z_{m},$ all elements of $Z_{m}$ having an inverse
must belong to $X$ . It is also obvious that $X$ is closed (with respect to
multiplication).\newline

3) Since it is evident that $\left( x+y\cdot z\right) \ $cannot belong to $Y$%
, then it must belong to $X$ (again from $X\cap Y=0,~X\cup Y=Z_{m}$)\medskip
\end{proof}

Let us now parametrize the field $u$ in the following way:%
\begin{equation}
u(n,t)=c_{1}+c_{2}~q(n,t),  \label{par}
\end{equation}%
where $c_{1},~c_{2}$ are (up to now) arbitrary constants.

As a consequence of the above \emph{Theorem}, we can project directly the
BTs (\ref{ca2}) in the finite ring $Z_{m},m=p^{k},~k\in 
%TCIMACRO{\U{2115} }%
%BeginExpansion
\mathbb{N}
%EndExpansion
,~p=prime$, provided that we choose $c_{1}\in X,~c_{2}\in Y$ . However, due
to (\ref{ds3}), we have to assume:%
\begin{equation}
c_{1}=1,~q(n,t)\underset{\left\vert n\right\vert \longrightarrow \infty }{%
\longrightarrow }0~.
\end{equation}%
In terms of the new field $q$, we obtain from (\ref{Ca2a}) the new cellular
automaton CA2a:%
\begin{equation*}
q(n,t+1)\overset{m}{=}q(n,t)+
\end{equation*}%
\begin{equation}
+\left( \frac{q(n+1,t)-q(n-1,t+1)}{1+c_{2}~q(n-1,t+1)}\right) \left(
b+c_{2}\dsum\limits_{j=-\infty }^{n-1}\left( q(j,t+1)-q(j.t\right) \right) ,
\end{equation}%
and, from (\ref{ca2b}), the new cellular automaton CA2b

\begin{equation*}
q(n,t+1)\overset{m}{=}q(n,t)+
\end{equation*}%
\begin{equation}
+b\left( \frac{q(n+1,t)-q(n-1,t+1)}{1+c_{2}~q(n-1,t+1)}\right)
\dprod\limits_{j=-\infty }^{n-1}\frac{1+c_{2}~q(j+1,t)}{1+c_{2}~q(j-1,t+1)}.
\end{equation}%
Note that these two new cellular automata are again "filter" CA, however, if
the initial \textit{datum} is given on a compact support, they are
computable. Investigation of the dynamics of the CA introduced in this paper
is postponed to a subsequent paper.

Since CA1, CA2a, CA2b (as well as CA0) are related to the same linear
problem (\ref{qw},\ref{se}), they share the same infinite (countable) set of
constants of motion. This set was derived for CA0 in ref. \cite{noi}; here
we report just the first conserved quantities (in terms of the above
parametrization):%
\begin{equation}
K_{1}\overset{m}{=}\dsum\limits_{s=-\infty }^{+\infty }q(s,0),
\end{equation}%
\begin{equation}
K_{2}\overset{m}{=}\dsum\limits_{s=-\infty }^{+\infty
}q(s,0)\dsum\limits_{r=s+2}^{+\infty }q(r,0),
\end{equation}%
\begin{equation}
K_{3}\overset{m}{=}\dsum\limits_{s=-\infty }^{+\infty }q(s,0)\left(
c_{2}\dsum\limits_{r=s+2}^{+\infty }q(r,0)\dsum\limits_{l=r+2}^{+\infty
}q(l,0)-\dsum\limits_{r=s+3}^{+\infty }q(r,0)\right) .
\end{equation}%
\bigskip

\bigskip \textbf{Aknowledgements}

\textit{We acknowledge useful discussions with A.K. Pogrebkov. It is also a
pleasure to thank P.M. Santini for discussions and for reading the
manuscript.}


\bigskip

\begin{thebibliography}{9}
\bibitem{wolfram} \bigskip S. Wolfram, "Theory and applications of cellular
automata", World scientific, Singapore, (1986)

\bibitem{filter} J.K.Park, K. Steiglitz and W.P. Thurston, Physica D19
(1986),423.

\bibitem{noi} M. Bruschi, P.M. Santini and O. Ragnisco, Phys. Lett. A169
(1992), 151.

\bibitem{kac} M.Kac and P. van Moerbeke, J. Math. Phys. 72 (1974), 2879;
S.V. Manakov, Sov. Phys. JTP 40 (1975), 269; M. Toda, "Theory of nonlinear
lattices", Springer, Berlin, (1981).
\end{thebibliography}
\end{document}